\documentclass[11pt]{article}
\usepackage{acl2013}
\usepackage{times}
\usepackage{caption,comment,graphicx,latexsym,multirow,subcaption,url}

\title{Complexity of Word Collocation Networks:\\A Preliminary Structural Analysis}

\author{Shibamouli Lahiri\\
Computer Science and Engineering\\
University of North Texas\\
Denton, TX 76207, USA\\
{\tt shibamoulilahiri@my.unt.edu}}

\date{}

\begin{document}
\maketitle

\begin{abstract}
In this paper, we explore complex network properties of word collocation networks~\cite{Ferret:2002:UCT:1072228.1072261} from four different genres. Each document of a particular genre was converted into a network of words with word collocations as edges. We analyzed graphically and statistically how the \emph{global properties} of these networks varied across different genres, and among different network types within the same genre. Our results indicate that the distributions of network properties are visually similar but statistically apart across different genres, and interesting variations emerge when we consider different network types within a single genre. We further investigate how the global properties change as we add more and more collocation edges to the graph of one particular genre, and observe that except for the number of vertices and the size of the largest connected component, network properties change in \emph{phases}, via jumps and drops.
\end{abstract}

\section{Introduction}
\label{sec:intro}

Word collocation networks~\cite{Ferret:2002:UCT:1072228.1072261,DBLP:journals/corr/abs-cs-0701135}, also known as collocation graphs~\cite{heyer2001learning,choudhury2009structure}, are networks of words found in a document or a document collection, where each node corresponds to a unique \emph{word type}, and edges correspond to \emph{word collocations}~\cite{DBLP:journals/jql/KeY08}. In the simplest case, each edge corresponds to a unique bigram in the original document. For example, if the words $w_A$ and $w_B$ appeared together in a document as a bigram $w_{A}w_{B}$, then the word collocation network of that particular document will contain an edge $w_{A}\rightarrow w_{B}$. Note that edges can be directed ($w_{A}\rightarrow w_{B}$) or undirected ($w_{A} - w_{B}$). Furthermore, they can be weighted (with the frequency of the bigram $w_{A}w_{B}$) or unweighted.

It is interesting to note that word collocation networks display complex network structure, including power-law degree distribution and small-world behavior~\cite{Matsuo:2001:DSW:645654.665684,Matsuo:2001:KEK:647858.738697,citeulike:4919958,study_on_co_occurrence}. This is not surprising, given that natural language generally shows complex network properties at different levels~\cite{canchosole_2001,MDLD02,DBLP:conf/acl/BiemannCM09,Liang20094901}. Moreover, researchers have used such complex networks in applications ranging from text genre identification~\cite{DBLP:journals/corr/abs-1007-3254} and Web query analysis~\cite{roy2011complex} to semantic analysis~\cite{coling2012_semantics_cn} and opinion mining~\cite{opinion_cn_recent}. In Section~\ref{sec:related_work}, we will discuss some of these applications in more detail.

The goal of this paper is to explore some key structural properties of these complex networks (cf. Table~\ref{tab:global_network_properties}), and study how they vary across different genres of text, and also across different network types within the same genre. We chose \emph{global network properties} like diameter, global clustering coefficient, shrinkage exponent~\cite{Leskovec:2007:GED:1217299.1217301}, and small-worldliness~\cite{conf/ijcai/Walsh99,Matsuo:2001:DSW:645654.665684}, and experimented with four different text collections -- blogs, news articles, academic papers, and digitized books (Section~\ref{subsec:datasets}). Six different types of word collocation networks were constructed on each document, as well as on the entire collections -- two with directed edges, and four with undirected edges (Section~\ref{sec:collocation_networks}). We did not take into account edge weights in our study, and kept it as a part of our future work (Section~\ref{sec:conclusion}).

Tracking the variation of complex network properties on word collocation networks yielded several important observations and insights. We noted in particular that different genres had considerable visual overlap in the distributions of global network properties like diameter and clustering coefficient (cf. Figure~\ref{fig:digraph_distributions}), although statistical significance tests indicated the distributions were sufficiently apart from each other (Section~\ref{subsec:distributions}). This calls for a deeper analysis of complex network properties and their general applicability to tasks like genre identification~\cite{DBLP:journals/corr/abs-1007-3254}.

We further analyzed distributions of global word network properties across six different network types \emph{within the same genre} (Section~\ref{subsec:distributions}). This time, however, we noted a significant amount of separation -- both visually as well as statistically -- among the distributions of different global properties (cf. Figure~\ref{fig:Reuters_distributions} and Table~\ref{tab:anova_kw}).

In our final set of experiments, we analyzed how global network properties change as we start with an empty network, and gradually add edges to that network. For this experiment, we chose the news genre, and tracked the variation of 17 different global network properties on four types of networks. We observed that all global network properties (except the number of vertices and edges, number of connected components and the size of the largest connected component) show unpredictability and \emph{spikes} when the percentage of added edges is small. We also noted that most global properties showed at least one \emph{phase transition} as the word collocation networks grew larger. Statistical significance tests indicated that the patterns of most global property variations were non-random and positively correlated (Section~\ref{subsec:incremental}).

\section{Related Work}
\label{sec:related_work}

\begin{figure*}
\centering
 \begin{subfigure}[b]{0.2\textwidth}
 \centering
 \includegraphics[width=\textwidth]{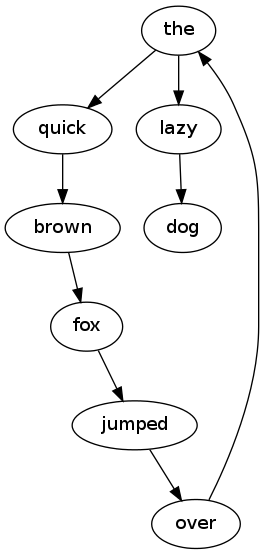}
 \caption{Directed}
 \label{fig:digraph}
 \end{subfigure}
~~~~~~~~~~~~~~~~~~
 \begin{subfigure}[b]{0.2\textwidth}
 \centering
 \includegraphics[width=\textwidth]{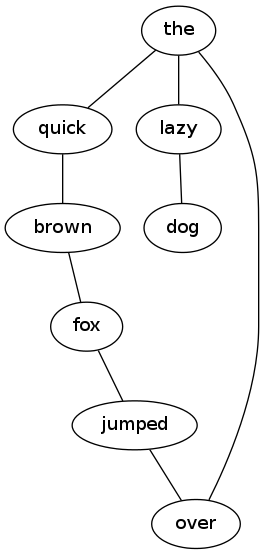}
 \caption{Undirected, Variant 1}
 \label{fig:undigraph1}
 \end{subfigure}
~~~~~~~~~~~~~~~~~~
 \begin{subfigure}[b]{0.2\textwidth}
 \centering
 \includegraphics[width=\textwidth]{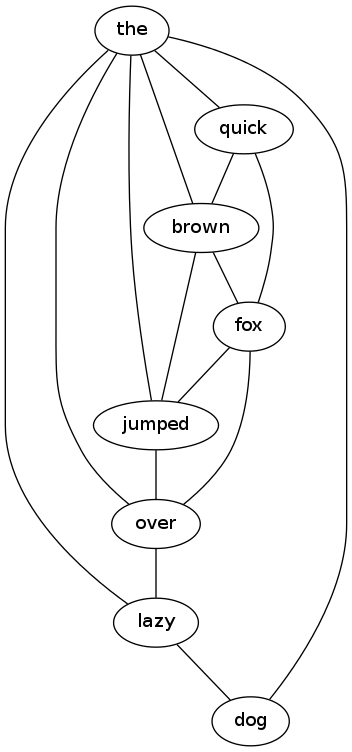}
 \caption{Undirected, Variant 2}
 \label{fig:undigraph2}
 \end{subfigure}
\caption{Word collocation networks of the sentence \emph{``the quick brown fox jumped over the lazy dog''}. Note that for all three network types, the word ``the'' appeared as the most central word. It is in general the case that stop words like ``the'' are the most central words in collocation networks, especially since they act as \emph{connectors} between other words.}
\label{fig:variants_of_word_networks}
\end{figure*}

\begin{table*}
\begin{center}
\begin{tabular}{lc}
\hline
\textbf{Network Property} & \textbf{Mathematical Expression}\\
\hline
Number of vertices & $|V|$\\
Number of edges & $|E|$\\
Shrinkage exponent~\cite{Leskovec:2007:GED:1217299.1217301} & $\log_{|V|}|E|$\\
Global clustering coefficient & $C$\\
Small-worldliness~\cite{conf/ijcai/Walsh99,Matsuo:2001:DSW:645654.665684} & $\mu = (\bar{C}/L)/(\bar{C}_{rand}/L_{rand})$\\
Diameter (directed) & $d$\\
Diameter (undirected) & $d$\\
Power-law exponent of degree distribution & $\alpha$\\
Power-law exponent of in-degree distribution & $\alpha_{in}$\\
Power-law exponent of out-degree distribution & $\alpha_{out}$\\
p-value for the power-law exponent of degree distribution & N/A\\
p-value for the power-law exponent of in-degree distribution & N/A\\
p-value for the power-law exponent of out-degree distribution & N/A\\
Number of connected components* & N/A\\
Size of the largest connected component* & N/A\\
Number of strongly connected components* & N/A\\
Size of the largest strongly connected component* & N/A\\
\hline
\end{tabular}
\end{center}
\caption{\label{tab:global_network_properties}Different global network properties used in our study. The ones marked with an asterisk (``*'') are only used in Section~\ref{subsec:incremental} in the context of incrementally constructing networks by gradually adding edges. For document networks, these four properties do not make sense, because the number of connected components is always one, and the size of the largest connected component always equals the number of vertices in the document network. Note also that in-degree distribution, out-degree distribution, and the directed version of diameter do not make sense for undirected networks, and same goes with the number of strongly connected components and the size of the largest strongly connected component. Here we report them separately for conceptual clarity.}
\end{table*}

That language shows complex network structure at the word level, was shown more than a decade ago by at least two independent groups of researchers~\cite{canchosole_2001,Matsuo:2001:DSW:645654.665684}. Matsuo et al.~\shortcite{Matsuo:2001:KEK:647858.738697} went further ahead, and designed an unsupervised keyword extraction algorithm using the small-world property of word collocation networks. Motter et al.~\shortcite{MDLD02} extended the collocation network idea to \emph{concepts} rather than words, and observed a small-world structure in the resulting network. Edges between concepts were defined as entries in an English thesaurus. Liang et al.~\shortcite{Liang20094901} compared word collocation networks of Chinese and English text, and pointed out their similarities and differences. They further constructed \emph{character collocation networks} in Chinese, showed their small-world structure, and used these networks in a follow-up study to accurately segregate Chinese essays from different literary periods~\cite{study_on_co_occurrence}.

Word collocation networks have also been successfully applied to the authorship attribution task.\footnote{For details on authorship attribution, please see the surveys by Juola~\shortcite{Juola:2006:AA:1373450.1373451}, Koppel et al.~\shortcite{Koppel:2009:CMA:1484611.1484627}, and Stamatatos~\shortcite{Stamatatos:2009:SMA:1527090.1527102}.} Antiqueira et al.~\shortcite{Antiqueira2006a} were among the first to apply complex network features like clustering coefficient, \emph{component dynamics deviation} and \emph{degree correlation} to the authorship attribution problem.

Biemann et al.~\shortcite{DBLP:conf/acl/BiemannCM09} constructed syntactic and semantic distributional similarity networks (DSNs), and analyzed their structural differences using spectral plots. Biemann et al.~\shortcite{coling2012_semantics_cn} further used \emph{graph motifs} on collocation networks to distinguish real natural language text from generated natural language text, and to point out the shortcomings of n-gram language models.

Word collocation networks have been used by Amancio et al.~\shortcite{opinion_cn_recent} for opinion mining, and by Mihalcea and Tarau~\shortcite{mihalcea-tarau:2004:EMNLP} for keyword extraction. While the former study used complex network properties as features for machine learning algorithms, the latter ran PageRank~\cite{Pageetal98} on word collocation networks to sieve out most important words.

While all the above studies are very important, we found none that performed a thorough and systematic exploration of different global network properties on different network types across genres, along with statistical significance tests to assess the validity of their observations. Stevanak et al.~\shortcite{DBLP:journals/corr/abs-1007-3254}, for example, used word collocation networks to distinguish between novels and news articles, but they did not perform a distributional analysis of the different global network properties they used, thereby leaving open how good those properties truly were as features for genre classification, and whether there exist a better and simpler set of global network properties for the same task. On the other hand, Masucci and Rodgers~\shortcite{citeulike:4919958}, Ke~\shortcite{DBLP:journals/corr/abs-cs-0701135}, and Ke and Yao~\shortcite{DBLP:journals/jql/KeY08} explored several global network properties on word collocation networks, but they did not address the problem of analyzing within-genre and cross-genre variations of those properties.

In addition to addressing these problems, in this paper we introduce a new analysis - how the global network properties change as we gradually add more collocation edges to a network (Section~\ref{subsec:incremental}).\footnote{All code, data, and supplementary material are available at \url{https://drive.google.com/file/d/0B2Mzhc7popBgODFKZVVnQTFMQkE/edit?usp=sharing}. The data includes -- among other things -- the corpora we used (cf. Section~\ref{subsec:datasets}), and code to construct the networks and analyze their properties.}

\section{Collocation Networks of Words}
\label{sec:collocation_networks}

Before constructing collocation networks, we lowercased the input text and removed all punctuation, but refrained from performing stemming in order to retain subtle distinctions between words like ``vector'' and ``vectorization''. Six different types of word collocation networks were constructed on each document (used in Section~\ref{subsec:distributions}) as well as on document collections (used in Section~\ref{subsec:incremental}), where nodes are unique words, and an edge appears between two nodes if their corresponding words appeared together as a bigram or in a trigram in the original text. All the network types have the \emph{same} number of vertices (i.e., words) for a particular document or a document collection, and they are only distinguished from each other by the type (and potentially, number) of edges, as follows:\\
\textbf{Directed} -- Directed edge $w_{A}\rightarrow w_{B}$ if $w_{A}w_{B}$ is a bigram in the given text.\\
\textbf{Undirected, Variant 1} -- Undirected edge $w_{A} - w_{B}$ if $w_{A}w_{B}$ is a bigram in the given text.\\
\textbf{Undirected, Variant 2} -- Undirected edges $w_{A} - w_{B}$, $w_{B} - w_{C}$ and $w_{A} - w_{C}$, if $w_{A}w_{B}w_{C}$ is a trigram in the given text.\\
\textbf{Directed Simplified} -- Same as the directed version, with \emph{self-loops} removed.\footnote{Note that self-loops may appear in word collocation networks due to punctuation removal in the pre-processing step. An example of such a self-loop is: \emph{``The airplane took off. Off we go to Alaska.''} Here the word ``off'' will contain a self-loop.}\\
\textbf{Undirected Variant 1, Simplified} -- Same as the undirected variant 1, with self-loops removed.\\
\textbf{Undirected Variant 2, Simplified} -- Same as the undirected variant 2, with self-loops removed.

We did not take into account edge weights in our study, and all our networks are therefore unweighted networks. Furthermore, since we removed all punctuation information \emph{before} constructing collocation networks, sentence boundaries were implicitly ignored. In other words, the last word of a sentence \emph{does} link to the first word of the next sentence in our collocation networks. An example of the first three types of networks (directed, undirected variant 1, and undirected variant 2) is shown in Figure~\ref{fig:variants_of_word_networks}. Here we considered a sentence \emph{``the quick brown fox jumped over the lazy dog''} as our document. Note that all the collocation networks in Figure~\ref{fig:variants_of_word_networks} contain at least one cycle, and the directed version contains a directed cycle. In a realistic document network, there can be many such cycles.

We constructed word collocation networks on \emph{document collections} as well. In this case, the six network types remain as before, and the only difference comes from the fact that now the whole collection is considered a single \emph{super-document}. Words in this super-document are connected according to bigram and trigram relationships. We respected document boundaries in this case, so the last word of a particular document \emph{does not} link to the first word of the next document. The \emph{collection networks} have only been used in Section~\ref{subsec:incremental} of this paper, to show how global network properties change as we add edges to the network.

\begin{figure*}
\centering
 \begin{subfigure}[b]{0.245\textwidth}
 \centering
 \includegraphics[width=\textwidth]{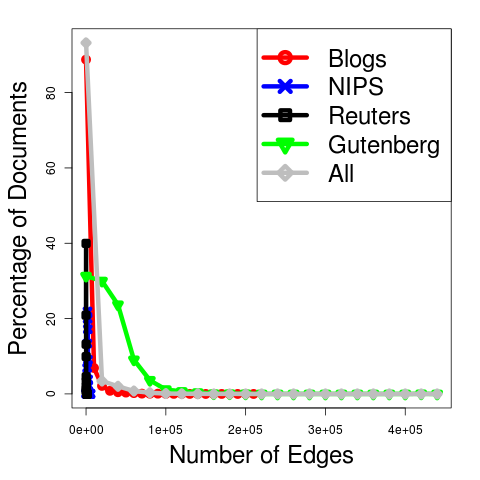}
 \caption{Number of Edges}
 \label{fig:digraph_number_of_edges}
 \end{subfigure}
 \begin{subfigure}[b]{0.245\textwidth}
 \centering
 \includegraphics[width=\textwidth]{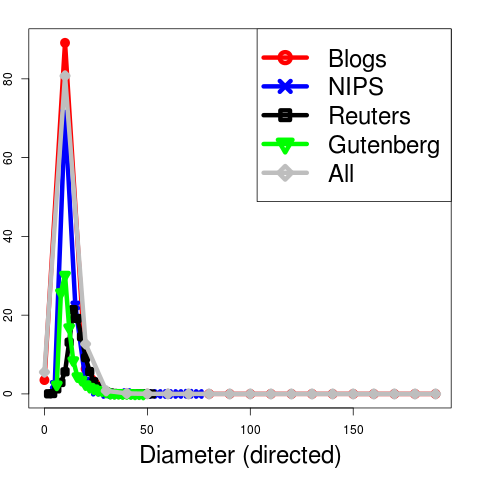}
 \caption{Diameter (directed)}
 \label{fig:digraph_directed_diameter}
 \end{subfigure}
 \begin{subfigure}[b]{0.245\textwidth}
 \centering
 \includegraphics[width=\textwidth]{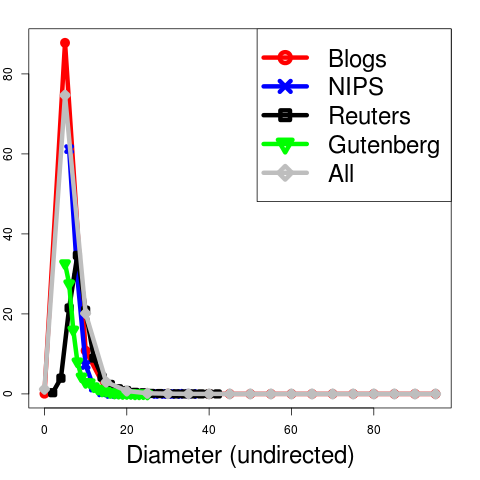}
 \caption{Diameter (undirected)}
 \label{fig:digraph_undirected_diameter}
 \end{subfigure}
 \begin{subfigure}[b]{0.245\textwidth}
 \centering
 \includegraphics[width=\textwidth]{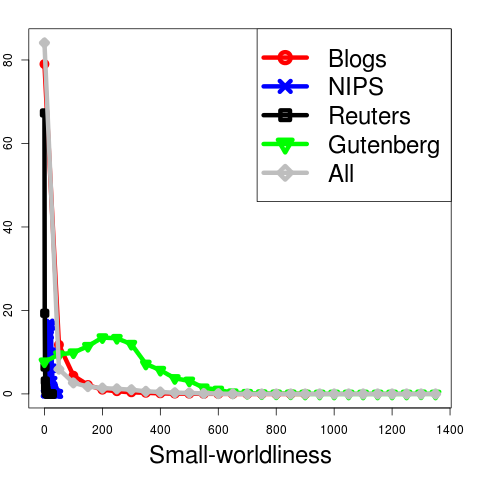}
 \caption{Small-worldliness}
 \label{fig:digraph_small_worldliness}
 \end{subfigure}

 \begin{subfigure}[b]{0.245\textwidth}
 \centering
 \includegraphics[width=\textwidth]{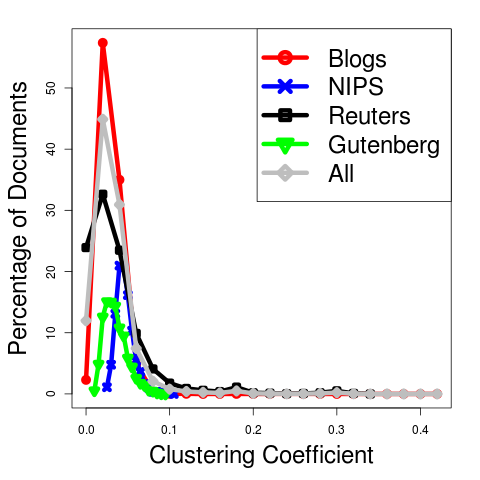}
 \caption{Clustering Coefficient}
 \label{fig:digraph_global_clustering_coefficient}
 \end{subfigure}
 \begin{subfigure}[b]{0.245\textwidth}
 \centering
 \includegraphics[width=\textwidth]{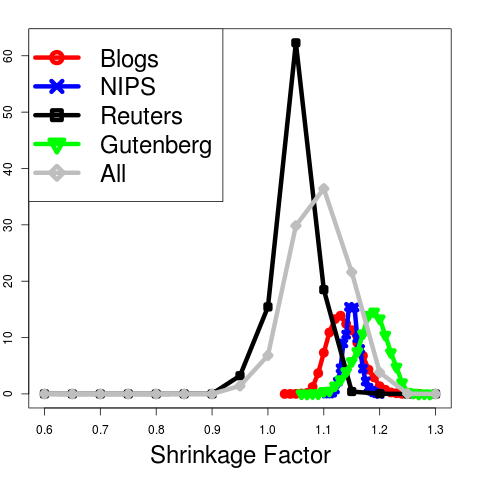}
 \caption{Shrinkage Exponent}
 \label{fig:digraph_leskovec_shrinkage_exponent}
 \end{subfigure}
 \begin{subfigure}[b]{0.245\textwidth}
 \centering
 \includegraphics[width=\textwidth]{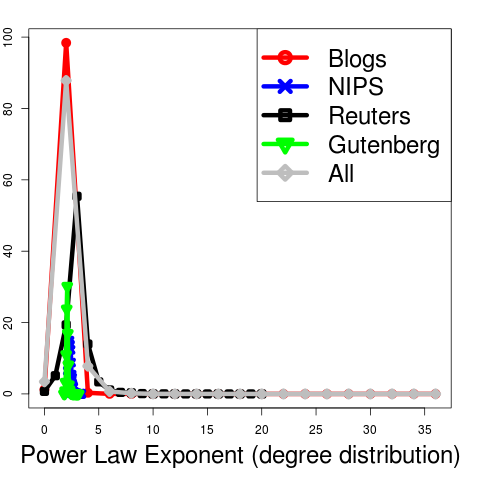}
 \caption{$\alpha$}
 \label{fig:digraph_power_law_exponent_all_degree_distribution}
 \end{subfigure}
 \begin{subfigure}[b]{0.245\textwidth}
 \centering
 \includegraphics[width=\textwidth]{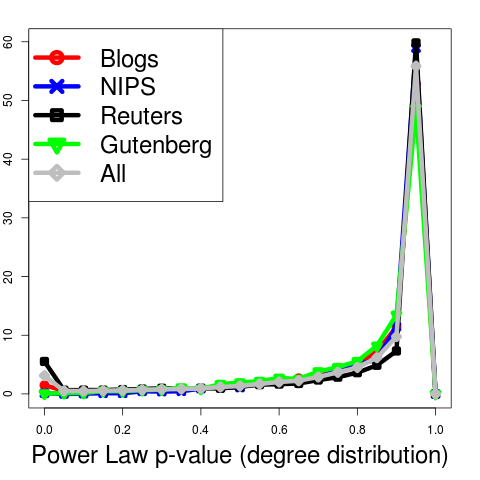}
 \caption{p-value for $\alpha$}
 \label{fig:digraph_power_law_pvalue_all_degree_distribution}
 \end{subfigure}
\caption{Distributions of eight global network properties across different genres for directed collocation networks. Y-axes represent the percentage of documents for different genres.}
\label{fig:digraph_distributions}
\end{figure*}

\begin{figure*}
\centering
 \begin{subfigure}[b]{0.245\textwidth}
 \centering
 \includegraphics[width=\textwidth]{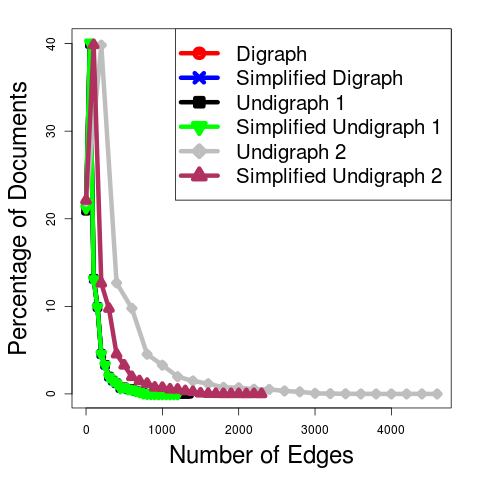}
 \caption{Number of Edges}
 \label{fig:Reuters_number_of_edges}
 \end{subfigure}
 \begin{subfigure}[b]{0.245\textwidth}
 \centering
 \includegraphics[width=\textwidth]{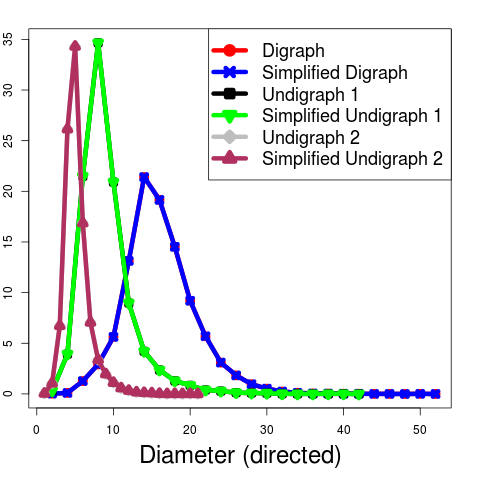}
 \caption{Diameter (directed)}
 \label{fig:Reuters_directed_diameter}
 \end{subfigure}
 \begin{subfigure}[b]{0.245\textwidth}
 \centering
 \includegraphics[width=\textwidth]{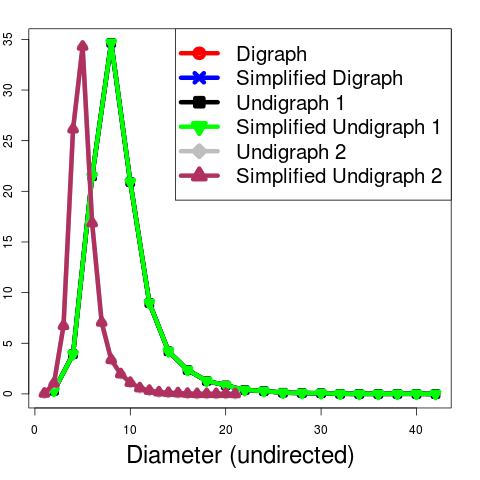}
 \caption{Diameter (undirected)}
 \label{fig:Reuters_undirected_diameter}
 \end{subfigure}
 \begin{subfigure}[b]{0.245\textwidth}
 \centering
 \includegraphics[width=\textwidth]{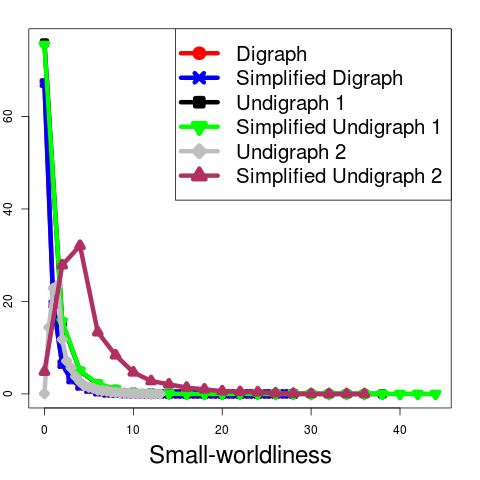}
 \caption{Small-worldliness}
 \label{fig:Reuters_small_worldliness}
 \end{subfigure}

 \begin{subfigure}[b]{0.245\textwidth}
 \centering
 \includegraphics[width=\textwidth]{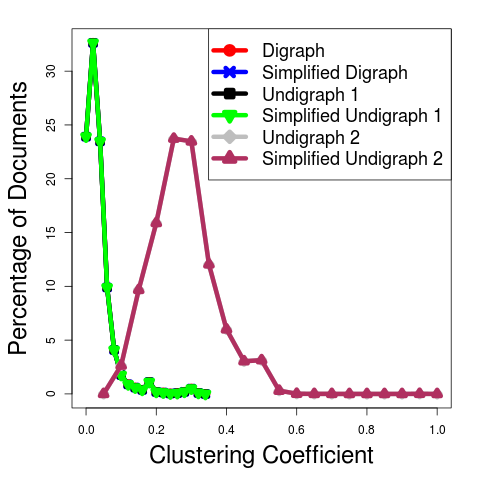}
 \caption{Clustering Coefficient}
 \label{fig:Reuters_global_clustering_coefficient}
 \end{subfigure}
 \begin{subfigure}[b]{0.245\textwidth}
 \centering
 \includegraphics[width=\textwidth]{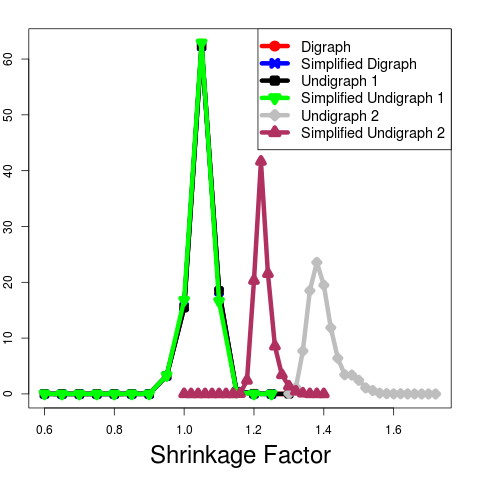}
 \caption{Shrinkage Exponent}
 \label{fig:Reuters_leskovec_shrinkage_exponent}
 \end{subfigure}
 \begin{subfigure}[b]{0.245\textwidth}
 \centering
 \includegraphics[width=\textwidth]{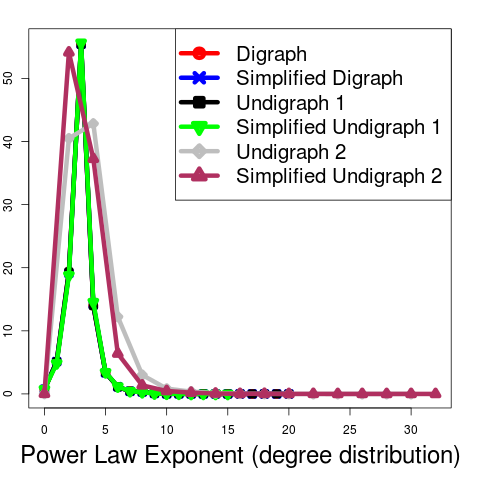}
 \caption{$\alpha$}
 \label{fig:Reuters_power_law_exponent_all_degree_distribution}
 \end{subfigure}
 \begin{subfigure}[b]{0.245\textwidth}
 \centering
 \includegraphics[width=\textwidth]{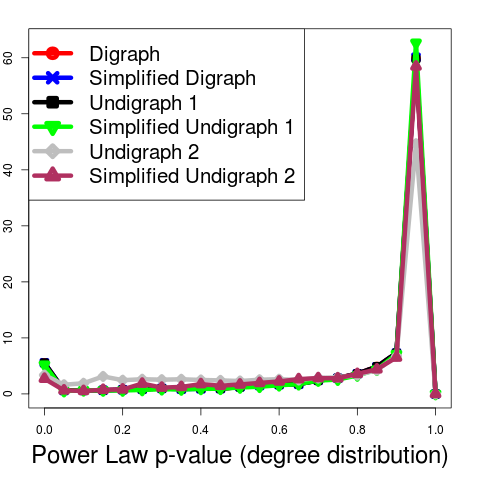}
 \caption{p-value for $\alpha$}
 \label{fig:Reuters_power_law_pvalue_all_degree_distribution}
 \end{subfigure}
\caption{Distributions of eight global network properties across different network types on the \emph{news} genre. Y-axes represent the percentage of documents for different network types.}
\label{fig:Reuters_distributions}
\end{figure*}

With the networks now constructed, we went ahead and explored several of their global properties (cf. Table~\ref{tab:global_network_properties}). Properties were measured on each type of network on each document, thereby giving us property distributions across different genres of documents for a particular network type (cf. Figure~\ref{fig:digraph_distributions}), as well as property distributions across different network types for a particular genre (cf. Figure~\ref{fig:Reuters_distributions}). We used the \emph{igraph} software package~\cite{igraph_cite} for computing global network properties.

Among the properties in Table~\ref{tab:global_network_properties}, number of vertices ($|V|$) and number of edges ($|E|$) are self-explanatory. The \emph{shrinkage exponent} ($\log_{|V|}|E|$) is motivated by the observations that the number of edges ($|E|$) follows a power-law relationship with the number of vertices ($|V|$), and that as a network evolves, both $|V|$ and $|E|$ continue to grow, but the diameter of the network either \emph{shrinks} or plateaus out, thereby resulting in a \emph{densified} network~\cite{Leskovec:2007:GED:1217299.1217301}. We explored two versions of graph diameter ($d$) in our study - a directed version (considering directed edges), and an undirected version (ignoring edge directions).\footnote{For undirected collocation networks, these two versions yield the same results, as expected.}

The \emph{global clustering coefficient} ($C$) is a measure of how interconnected a graph's nodes are among themselves. It is defined as the ratio between the number of closed triplets of vertices (i.e., the number of ordered triangles or \emph{transitive triads}), and the number of connected vertex-triples~\cite{Wasserman1994}. The \emph{small-worldliness} or \emph{proximity ratio} ($\mu$) of a network measures to what extent the network exhibits small-world behavior. It is quantified as the amount of deviation of the network from an equally large random network, in terms of average local clustering coefficient ($\bar{C}$) and average shortest path length ($L$)\footnote{Also called \emph{``characteristic path length''}~\cite{watts1998cds}.}. The exact ratio is $\mu = (\bar{C}/L)/(\bar{C}_{rand}/L_{rand})$, where $\bar{C}$ and $L$ are the average local clustering coefficient and the average shortest path length of the given network, and $\bar{C}_{rand}$ and $L_{rand}$ are the average local clustering coefficient and the average shortest path length of an equally large random network~\cite{conf/ijcai/Walsh99,Matsuo:2001:DSW:645654.665684}.

Since collocation networks have been found to display scale-free (power-law) degree distribution in several previous studies (see, e.g.,~\cite{canchosole_2001,citeulike:4919958,Liang20094901}), we computed power-law exponents of in-degree, out-degree, and degree distributions on each of our collocation networks.\footnote{For undirected graphs, the exponents on all three distributions are the same.} We also computed the corresponding p-values, following a procedure outlined in~\cite{Clauset:2009:PDE:1655787.1655789}. These p-values help assess whether the distributions are power-law or not. If a p-value is $< 0.05$, then there is statistical evidence to believe that the corresponding distribution is \emph{not} a power-law distribution.

Finally, we computed the number of connected components, size of the largest (``giant'') connected component, number of strongly connected components, and size of the largest strongly connected component, to be used in Section~\ref{subsec:incremental}.

\section{Analysis of Network Properties}
\label{sec:analysis_of_nw_properties}

\begin{table*}
\begin{center}
\footnotesize
\begin{tabular}{lccc|ccc}
\hline
\multirow{4}{*}{\textbf{Dataset}} & \multirow{2}{*}{\textbf{Median $\alpha$}} & \multirow{2}{*}{\textbf{Median $\alpha$}} & \multirow{2}{*}{\textbf{Median $\alpha$}} & \multirow{2}{*}{\textbf{Median $\mu$}} & \multirow{2}{*}{\textbf{Median $\mu$}} & \multirow{2}{*}{\textbf{Median $\mu$}}\\
& \multirow{2}{*}{\textbf{on Digraph}} & \multirow{2}{*}{\textbf{on Undigraph 1}} & \multirow{2}{*}{\textbf{on Undigraph 2}} & \multirow{2}{*}{\textbf{on Digraph}} & \multirow{2}{*}{\textbf{on Undigraph 1}} & \multirow{2}{*}{\textbf{on Undigraph 2}}\\
&&&&&&\\
& \multicolumn{3}{c|}{(quartile deviations are in parentheses)} & \multicolumn{3}{c}{(quartile deviations are in parentheses)}\\
\hline
Blog & 2.34 (0.17) & 2.34 (0.17) & 2.41 (0.19) & 16.63 (17.16) & 22.50 (22.01) & 14.93 (9.49)\\
News & 3.38 (0.42) & 3.38 (0.42) & 4.35 (0.98) & 0.63 (0.50) & 0.95 (0.76) & 1.75 (0.71)\\
Papers & 2.35 (0.09) & 2.35 (0.09) & 2.45 (0.11) & 20.69 (2.96) & 27.87 (3.93) & 14.95 (1.80)\\
Digitized Books & 2.12 (0.04) & 2.12 (0.04) & 2.16 (0.05) & 244.31 (98.62) & 296.73 (116.98) & 88.46 (31.78)\\
All together & 2.58 (0.53) & 2.58 (0.53) & 2.70 (0.90) & 5.03 (11.93) & 7.27 (15.85) & 7.31 (8.47)\\
\hline
\end{tabular}
\end{center}
\caption{\label{tab:median_alpha_mu_for_different_datasets} Power-law exponent of degree distribution ($\alpha$) and small-worldliness ($\mu$) of word collocation networks. Here we report the median across documents in a particular dataset (genre), and also the median across all documents in all datasets (last row).
}
\end{table*}

\begin{table*}
\begin{center}
\tiny
\begin{tabular}{lccc|ccc}
\hline
\multirow{4}{*}{\textbf{Dataset}} & \multirow{2}{*}{\textbf{Median $\alpha$} on} & \multirow{2}{*}{\textbf{Median $\alpha$} on} & \multirow{2}{*}{\textbf{Median $\alpha$} on} & \multirow{2}{*}{\textbf{Median $\mu$} on} & \multirow{2}{*}{\textbf{Median $\mu$} on} & \multirow{2}{*}{\textbf{Median $\mu$} on}\\
& \multirow{2}{*}{\textbf{Simplified Digraph}} & \multirow{2}{*}{\textbf{Simplified Undigraph 1}} & \multirow{2}{*}{\textbf{Simplified Undigraph 2}} & \multirow{2}{*}{\textbf{Simplified Digraph}} & \multirow{2}{*}{\textbf{Simplified Undigraph 1}} & \multirow{2}{*}{\textbf{Simplified Undigraph 2}}\\
&&&&&&\\
& \multicolumn{3}{c|}{(quartile deviations are in parentheses)} & \multicolumn{3}{c}{(quartile deviations are in parentheses)}\\
\hline
Blog & 2.34 (0.17) & 2.34 (0.16) & 2.36 (0.18) & 16.67 (17.18) & 23.28 (22.98) & 39.13 (24.03)\\
News & 3.39 (0.42) & 3.40 (0.42) & 3.88 (0.77) & 0.63 (0.50) & 0.96 (0.77) & 4.96 (1.93)\\
Papers & 2.36 (0.09) & 2.37 (0.09) & 2.40 (0.11) & 20.78 (2.98) & 29.18 (4.09) & 38.81 (4.75)\\
Digitized Books & 2.12 (0.04) & 2.13 (0.04) & 2.14 (0.05) & 244.53 (98.81) & 317.49 (127.14) & 218.77 (78.02)\\
All together & 2.58 (0.53) & 2.58 (0.54) & 2.65 (0.72) & 5.04 (11.97) & 7.45 (16.52) & 19.64 (21.82)\\
\hline
\end{tabular}
\end{center}
\caption{\label{tab:median_alpha_mu_for_different_datasets_2} Power-law exponent of degree distribution ($\alpha$) and small-worldliness ($\mu$) of word collocation networks. Here we report the median across documents in a particular dataset (genre), and also the median across all documents in all datasets (last row).
}
\end{table*}

\begin{table*}
\begin{center}
\tiny
\begin{tabular}{lccccc|ccccc}
\hline
\multirow{4}{*}{\textbf{Network Type}} & \multirow{2}{*}{\textbf{Median $\alpha$}} & \multirow{2}{*}{\textbf{Median $\alpha$}} & \multirow{2}{*}{\textbf{Median $\alpha$}} & \multirow{2}{*}{\textbf{Median $\alpha$}} & \multirow{2}{*}{\textbf{Median $\alpha$}} & \multirow{2}{*}{\textbf{Median $\mu$}} & \multirow{2}{*}{\textbf{Median $\mu$}} & \multirow{2}{*}{\textbf{Median $\mu$}} & \multirow{2}{*}{\textbf{Median $\mu$}} & \multirow{2}{*}{\textbf{Median $\mu$}}\\
& \multirow{2}{*}{\textbf{on Blogs}} & \multirow{2}{*}{\textbf{on Papers}} & \multirow{2}{*}{\textbf{on News}} & \multirow{2}{*}{\textbf{on Books}} & \multirow{2}{*}{\textbf{on All}} & \multirow{2}{*}{\textbf{on Blogs}} & \multirow{2}{*}{\textbf{on Papers}} & \multirow{2}{*}{\textbf{on News}} & \multirow{2}{*}{\textbf{on Books}} & \multirow{2}{*}{\textbf{on All}}\\
&&&&&&&&&&\\
& \multicolumn{5}{c|}{(quartile deviations are in parentheses)} & \multicolumn{5}{c}{(quartile deviations are in parentheses)}\\
\hline
Digraph & 2.34 (0.17) & 2.35 (0.09) & 3.38 (0.42) & 2.12 (0.04) & 2.58 (0.53) & 16.63 (17.16) & 20.69 (2.96) & 0.63 (0.50) & 244.31 (98.62) & 5.03 (11.93)\\
Undigraph 1 & 2.34 (0.17) & 2.35 (0.09) & 3.38 (0.42) & 2.12 (0.04) & 2.58 (0.53) & 22.50 (22.01) & 27.87 (3.93) & 0.95 (0.76) & 296.73 (116.98) & 7.27 (15.85)\\
Undigraph 2 & 2.41 (0.19) & 2.45 (0.11) & 4.35 (0.98) & 2.16 (0.05) & 2.70 (0.90) & 14.93 (9.49) & 14.95 (1.80) & 1.75 (0.71) & 88.46 (31.78) & 7.31 (8.47)\\
Simplified Digraph & 2.34 (0.17) & 2.36 (0.09) & 3.39 (0.42) & 2.12 (0.04) & 2.58 (0.53) & 16.67 (17.18) & 20.78 (2.98) & 0.63 (0.50) & 244.53 (98.81) & 5.04 (11.97)\\
Simplified Undigraph 1 & 2.34 (0.16) & 2.37 (0.09) & 3.40 (0.42) & 2.13 (0.04) & 2.58 (0.54) & 23.28 (22.98) & 29.18 (4.09) & 0.96 (0.77) & 317.49 (127.14) & 7.45 (16.52)\\
Simplified Undigraph 2 & 2.36 (0.18) & 2.40 (0.11) & 3.88 (0.77) & 2.14 (0.05) & 2.65 (0.72) & 39.13 (24.03) & 38.81 (4.75) & 4.96 (1.93) & 218.77 (78.02) & 19.64 (21.82)\\
\hline
\end{tabular}
\end{center}
\caption{\label{tab:median_alpha_mu_for_different_network_types} Power-law exponent of degree distribution ($\alpha$) and small-worldliness ($\mu$) of word collocation networks. Here we report the median across documents for a particular network type.
}
\end{table*}


\begin{table*}
\begin{center}
\footnotesize
\begin{tabular}{lcccccccc}
\hline
\textbf{Test} & \textbf{$|E|$} & \textbf{Directed $d$} & \textbf{Undirected $d$} & \textbf{$\mu$} & \textbf{$C$} & \textbf{Shrinkage} & \textbf{$\alpha$} & \textbf{p-value for $\alpha$}\\
\hline
ANOVA & $< 0.001$ & $< 0.001$ & $< 0.001$ & $< 0.001$ & $< 0.001$ & $< 0.001$ & $< 0.001$ & $< 0.001$ \\
Kruskal-Wallis & $< 0.001$ & $< 0.001$ & $< 0.001$ & $< 0.001$ & $< 0.001$ & $< 0.001$ & $< 0.001$ & $< 0.001$ \\
\hline
ANOVA & $< 0.001$ & $< 0.001$ & $< 0.001$ & $< 0.001$ & $< 0.001$ & $< 0.001$ & $< 0.001$ & $< 0.001$ \\
Kruskal-Wallis & $< 0.001$ & $< 0.001$ & $< 0.001$ & $< 0.001$ & $< 0.001$ & $< 0.001$ & $< 0.001$ & $< 0.001$ \\
\hline
\end{tabular}
\end{center}
\caption{\label{tab:anova_kw} p-values from ANOVA and Kruskal-Wallis tests. The top two rows are p-values for Figure~\ref{fig:digraph_distributions}, and the bottom two rows are p-values for Figure~\ref{fig:Reuters_distributions}. Each column corresponds to one subfigure of Figure~\ref{fig:digraph_distributions} and Figure~\ref{fig:Reuters_distributions}. p-values in general were extremely low - close to zero in most cases.
}
\end{table*}

\begin{figure*}
\centering
 \begin{subfigure}[b]{0.16\textwidth}
 \centering
 \includegraphics[width=\textwidth]{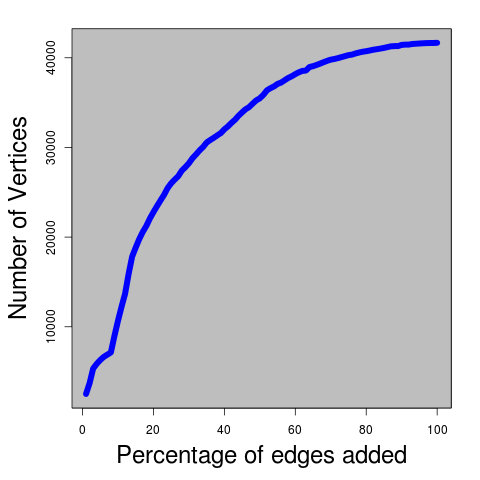}
 \caption{$|V|$}
 \label{fig:digraph_file_number_of_vertices}
 \end{subfigure}
 \begin{subfigure}[b]{0.16\textwidth}
 \centering
 \includegraphics[width=\textwidth]{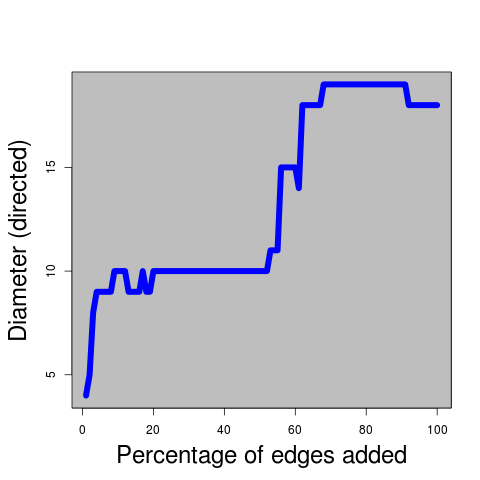}
 \caption{$d$ (directed)}
 \label{fig:digraph_file_directed_diameter}
 \end{subfigure}
 \begin{subfigure}[b]{0.16\textwidth}
 \centering
 \includegraphics[width=\textwidth]{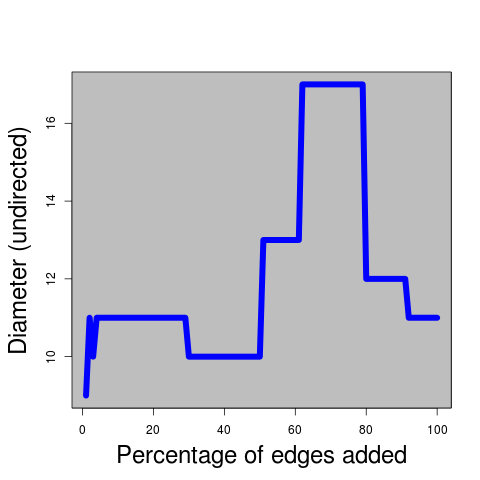}
 \caption{$d$ (undirected)}
 \label{fig:digraph_file_undirected_diameter}
 \end{subfigure}
 \begin{subfigure}[b]{0.16\textwidth}
 \centering
 \includegraphics[width=\textwidth]{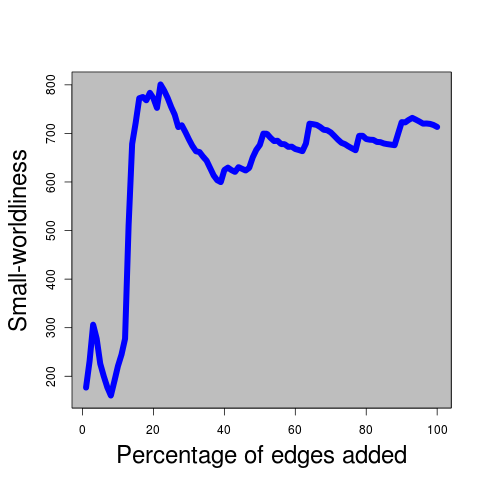}
 \caption{$\mu$}
 \label{fig:digraph_file_small_worldliness}
 \end{subfigure}
 \begin{subfigure}[b]{0.16\textwidth}
 \centering
 \includegraphics[width=\textwidth]{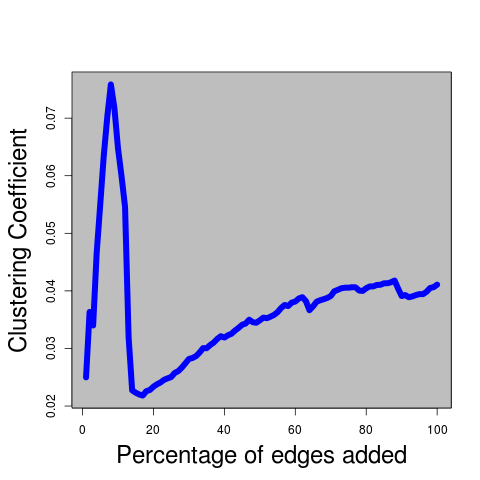}
 \caption{$C$}
 \label{fig:digraph_file_global_clustering_coefficient}
 \end{subfigure}
 \begin{subfigure}[b]{0.16\textwidth}
 \centering
 \includegraphics[width=\textwidth]{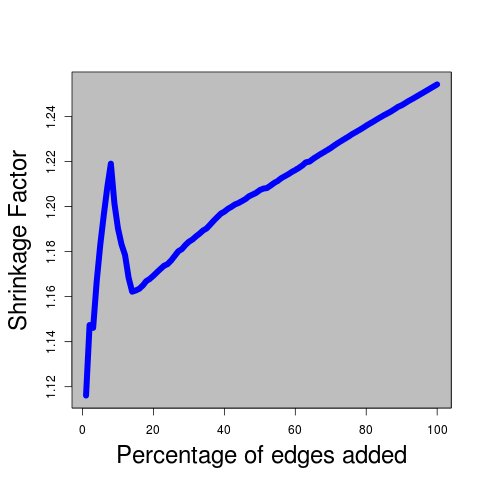}
 \caption{Shrinkage}
 \label{fig:digraph_file_leskovec_shrinkage_exponent}
 \end{subfigure}

 \begin{subfigure}[b]{0.16\textwidth}
 \centering
 \includegraphics[width=\textwidth]{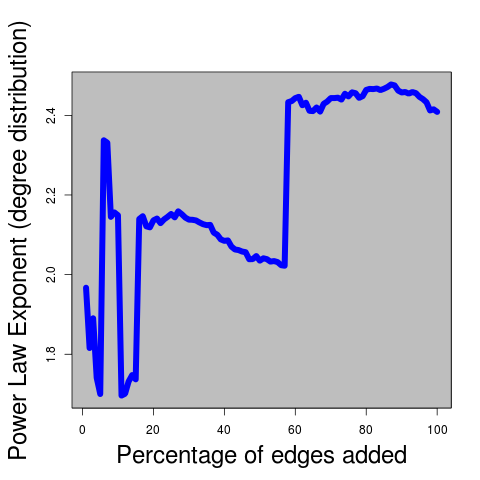}
 \caption{$\alpha$}
 \label{fig:digraph_file_power_law_exponent_all_degree_distribution}
 \end{subfigure}
 \begin{subfigure}[b]{0.16\textwidth}
 \centering
 \includegraphics[width=\textwidth]{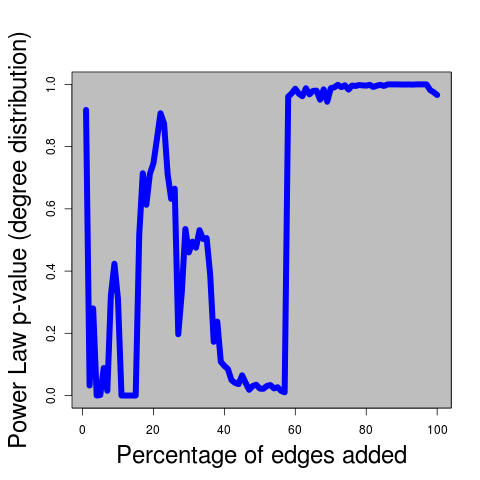}
 \caption{p-value for $\alpha$}
 \label{fig:digraph_file_power_law_pvalue_all_degree_distribution}
 \end{subfigure}
 \begin{subfigure}[b]{0.16\textwidth}
 \centering
 \includegraphics[width=\textwidth]{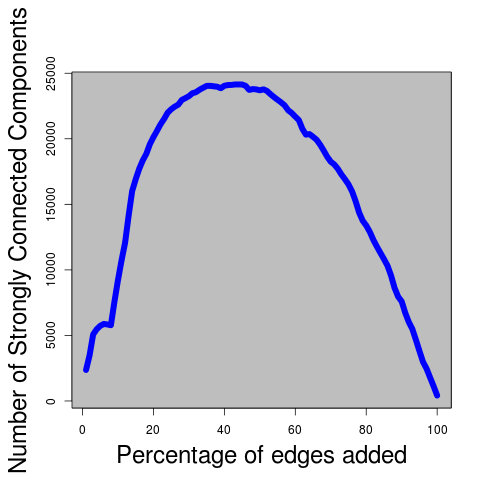}
 \caption{Number of SCCs}
 \label{fig:digraph_file_num_scc}
 \end{subfigure}
 \begin{subfigure}[b]{0.16\textwidth}
 \centering
 \includegraphics[width=\textwidth]{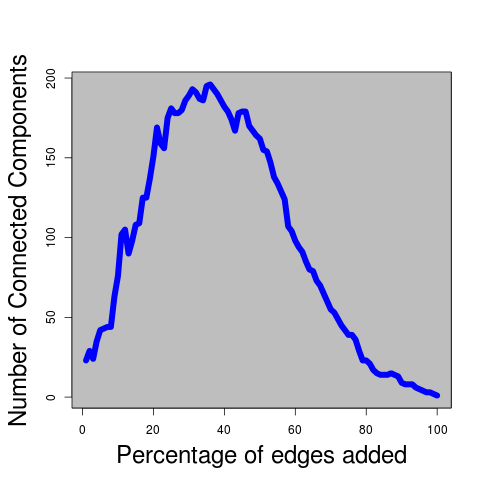}
 \caption{Number of CCs}
 \label{fig:digraph_file_num_wcc}
 \end{subfigure}
 \begin{subfigure}[b]{0.16\textwidth}
 \centering
 \includegraphics[width=\textwidth]{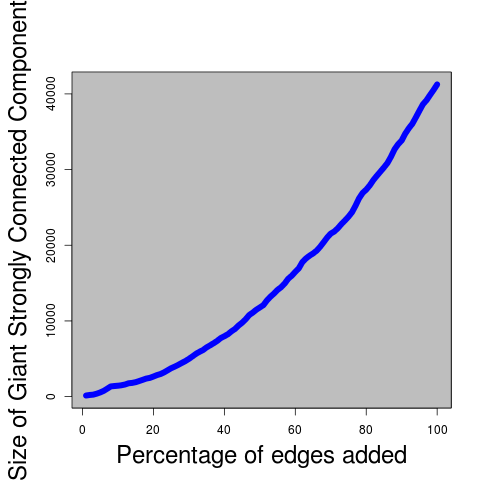}
 \caption{Giant SCC Size}
 \label{fig:digraph_file_size_of_giant_scc}
 \end{subfigure}
 \begin{subfigure}[b]{0.16\textwidth}
 \centering
 \includegraphics[width=\textwidth]{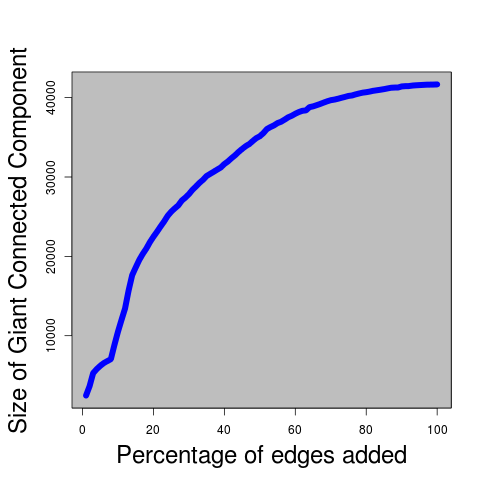}
 \caption{Giant CC Size}
 \label{fig:digraph_file_size_of_giant_wcc}
 \end{subfigure}
\caption{Change of global network properties with incremental addition of edges to the directed network of \emph{news} genre. SCC = Strongly Connected Component, CC = Connected Component. By ``giant'' CC and ``giant'' SCC, we mean the largest CC and the largest SCC. See Table~\ref{tab:global_network_properties} for other properties.}
\label{fig:digraph_file_incremental}
\end{figure*}

\subsection{Datasets}
\label{subsec:datasets}

We used four document collections from four different genres -- blogs, news articles, academic papers, and digitized books. For blogs, we used the \textbf{Blog Authorship Corpus} created by~\cite{koppel_2006_demog_blogging}. It consists of 19,320 blogs from authors of different age groups and professions. The unprocessed corpus has about 136.8 million word tokens.

Our news articles come from the \textbf{Reuters-21578, Distribution 1.0} collection.\footnote{Available from \url{http://www.daviddlewis.com/resources/testcollections/reuters21578/}.} This collection contains 19,043 news stories, and about 2.6 million word tokens (unprocessed).

For the academic paper dataset, we used \textbf{NIPS Conference Papers Vols 0-12}.\footnote{Available from \url{http://www.cs.nyu.edu/~roweis/data.html}.} This corpus comprises 1,740 papers and about 4.8 million unprocessed word tokens.

Finally, we created our own corpus of 3,036 digitized books written by 142 authors from the \textbf{Project Gutenberg} digital library.\footnote{\url{http://www.gutenberg.org/}.} After removing metadata, license information, and transcribers' notes, this dataset contains about 210.9 million word tokens.

That the word collocation networks of individual documents are indeed scale-free and small-world, is evident from Tables~\ref{tab:median_alpha_mu_for_different_datasets},~\ref{tab:median_alpha_mu_for_different_datasets_2}, and~\ref{tab:median_alpha_mu_for_different_network_types}, and Figure~\ref{fig:digraph_power_law_pvalue_all_degree_distribution}. Irrespective of network type, a majority of the median $\alpha$ (power-law exponent of degree distribution) values hovers in the range $[2,3]$, with low dispersion. This corroborates with earlier studies~\cite{canchosole_2001,Liang20094901,study_on_co_occurrence}. Similarly, the median $\mu$ (small-worldliness) is high for all genres except \emph{news} (irrespective of network type), thereby indicating the document networks are indeed small-world. This finding is in line with previous studies~\cite{Matsuo:2001:DSW:645654.665684,Matsuo:2001:KEK:647858.738697}. Moreover, Figure~\ref{fig:digraph_power_law_pvalue_all_degree_distribution} shows that a majority of documents in different genres have a very high p-value, indicating that the networks are significantly power-law. The \emph{news} genre poses an interesting case. Since many news stories in the Reuters-21578 collection are small, their collocation networks are not very well-connected, thereby resulting in very low small-worldliness values, as well as higher estimates of the power-law exponent $\alpha$ (cf. Tables~\ref{tab:median_alpha_mu_for_different_datasets},~\ref{tab:median_alpha_mu_for_different_datasets_2}, and~\ref{tab:median_alpha_mu_for_different_network_types}).

\subsection{Distribution of Global Network Properties}
\label{subsec:distributions}

We plotted the histograms of eight important global network properties on directed collocation networks in Figure~\ref{fig:digraph_distributions}. All histograms were plotted with 20 bins. Figure~\ref{fig:digraph_global_clustering_coefficient}, for example, shows the global clustering coefficient ($C$) on the X-axis, divided into 20 bins, and the percentage of document networks (directed) with $C$ values falling into a particular bin, on the Y-axis. Histograms from different genres are overlaid. Note from Figure~\ref{fig:digraph_global_clustering_coefficient} that most distributions are highly overlapping across different genres, thereby putting into question if they are indeed suitable for genre identification. But when we performed ANOVA and Kruskal-Wallis tests to figure out if the distributions were similar or not across different genres, we observed that the corresponding p-values were all $< 0.001$ (cf. Table~\ref{tab:anova_kw}, top two rows), thereby showing that at least a pair of mean values were significantly apart. Follow-up experiments using unpaired t-tests, U-tests, and Kolmogorov-Smirnov tests (all with Bonferroni Correction for multiple comparisons) showed that indeed almost all distributions across different genres were significantly apart from each other. Detailed results are in the supplementary material. This, we think, is an important and interesting finding, and needs to be delved deeper in future work.

Figure~\ref{fig:Reuters_distributions} shows histograms of the eight properties from Figure~\ref{fig:digraph_distributions}, but this time on a \emph{single genre} (news articles), across different network types. This time we observed that many histograms are significantly apart from each other (see, e.g., Figures~\ref{fig:Reuters_directed_diameter},~\ref{fig:Reuters_undirected_diameter},~\ref{fig:Reuters_global_clustering_coefficient}, and~\ref{fig:Reuters_leskovec_shrinkage_exponent}). ANOVA and Kruskal-Wallis tests corroborated this finding (cf. Table~\ref{tab:anova_kw}, bottom two rows). Detailed results, including t-tests, U-tests, and Kolmogorov-Smirnov tests are in the supplementary material.

\subsection{Change of Global Network Properties with Gradual Addition of Edges}
\label{subsec:incremental}

To see how global network properties change as we gradually add edges to a network, we took the whole news collection, and constructed a directed word collocation network on the whole collection, essentially considering the collection as a \emph{super-document} (cf. Section~\ref{sec:collocation_networks}). We studied how properties change as we consider top $k\%$ of edges in this super-network, with $k$ ranging from 1 to 100 in steps of 1. The result is shown in Figure~\ref{fig:digraph_file_incremental}. Note that the number of connected components and the number of strongly connected components increase first, and then decrease. The number of vertices, size of the largest strongly connected component, and size of the largest connected component increase monotonically as we consider more and more collocation edges. For other properties, we see a lot of unpredictability and spikes (see, e.g., Figures~\ref{fig:digraph_file_small_worldliness},~\ref{fig:digraph_file_global_clustering_coefficient},~\ref{fig:digraph_file_power_law_exponent_all_degree_distribution}, and~\ref{fig:digraph_file_power_law_pvalue_all_degree_distribution}), especially when the percentage of added edges is small. We performed Runs Test, Bartels Test, and Mann-Kendall Test to figure out if these trends are random, and the resulting p-values indicate that they are not random, and in fact positively correlated (i.e., \emph{increasing}). Details of these tests are in the supplementary material. Note also that all figures except Figures~\ref{fig:digraph_file_number_of_vertices},~\ref{fig:digraph_file_size_of_giant_scc}, and~\ref{fig:digraph_file_size_of_giant_wcc} show at least one \emph{phase transition} (i.e., a ``jump'' or a ``bend'').

\section{Conclusion}
\label{sec:conclusion}

We performed an exploratory analysis of global properties of word collocation networks across four different genres of text, and across different network types within the same genre. Our analyses reveal that cross-genre and within-genre variations are statistically significant, and incremental construction of collocation networks by gradually adding edges leads to non-random and positively correlated fluctuations in many global properties, some of them displaying single or multiple \emph{phase transitions}. Future work consists of the inclusion of edge weights; exploration of other datasets, network properties, and network types; and applying those properties to the genre classification task.

\section*{Acknowledgments}
\label{sec:acknowledgments}

We would like to acknowledge Dr Rada Mihalcea for her support. This work emerged from a class project in a graduate course on Network Science, given by Prof R\'{e}ka Albert at Penn State. Sagnik Ray Choudhury provided us with the primary inspiration to do the hard work and write the paper. Finally, we thank the anonymous reviewers whose comments greatly improved this draft. All results, discussions, and comments contained herein are the sole responsibility of the author, and in no way associated with any of the above-mentioned people. The errors and omissions, if any, should be addressed to the author, and will be gratefully acknowledged.

\bibliographystyle{acl}
\bibliography{forpaper}

\end{document}